\documentclass[10pt,a4paper]{article}
\usepackage[utf8]{inputenc}
\usepackage{float}
\usepackage{booktabs}
\usepackage{amsmath}
\usepackage{amssymb}
\usepackage{graphicx}

\makeatletter

\pdfpageheight\paperheight
\pdfpagewidth\paperwidth

\providecommand{\tabularnewline}{\\}

\usepackage{amsfonts}

\makeatother

\begin{document}
\title{Analyzing leptonic decays of heavy quarkonia in covariant confined
quark model}
\author{Stanislav Dubni\v{c}ka$^{1}$,\\
Anna Zuzana Dubni\v{c}kov\'{a}$^{2}$,\\
Mikhail A. Ivanov$^{3}$,\\
Andrej Liptaj$^{1\ddagger}$\\
Akmaral Tyulemissova$^{3,4}$\\
Zhomart Tyulemissov$^{3,4,5}$}
\maketitle
\begin{center}
$^{1}$ Institute of Physics, Slovak Academy of Sciences, Bratislava,
Slovakia.
\par\end{center}

\begin{center}
$^{2}$ Faculty of Mathematics, Physics and Informatics,\\
Comenius University, Bratislava, Slovakia.
\par\end{center}

\begin{center}
$^{3}$ Bogoliubov Laboratory of Theoretical Physics,\\
Joint Institute for Nuclear Research, Dubna, Russia.
\par\end{center}

\begin{center}
$^{4}$Institute of Nuclear Physics, Ministry of Energy\\
of the Republic of Kazakhstan, Almaty, Kazakhstan.
\par\end{center}

\begin{center}
$^{5}$Al-Farabi Kazakh National University\\
Almaty, Kazakhstan\\
\par\end{center}

\begin{center}
$^{\ddagger}$Andrej.Liptaj@savba.sk
\par\end{center}
\begin{abstract}
The leptonic decay widths of heavy quarkonia are complementary observables
to the semileptonic decays of heavy mesons and provide additional
insight into the possible lepton flavor universality breaking. We
analyze the former in the framework of the covariant confined quark
model, introducing a novel approach for the description of the radially
excited quarkonia states, which is based on constituent quark mass
running and the orthogonality of the states. We are able to describe
the experimental observations by our approach within the Standard
model, which does not support new physics explanations.
\end{abstract}


\section{Introduction}

The heavy quarkonia $J/\psi,\Psi(2s),\Upsilon(\text{n}s)_{\text{n}=1,2,3}$
are among the most simple realizations of quantum chromodynamics (QCD)
and naturally play a role in various physics areas (quark masses,
determination of $\alpha_{s}$, confinement, quark-gluon plasma and
QCD in general \cite{Kuhn:1998uy,Peset:2018ria,Resag:1994ki,Narison:2018dcr,QuarkoniumWorkingGroup:2004kpm,Brambilla:2010cs,Rothkopf:2019ipj}).
Here we study their leptonic decays which have the advantage of rather
clean experimental signatures and can be compared to theoretical predictions
at the level which has a discriminating power. This becomes useful
when the lepton flavor universality question is addressed, where interesting
developments are ongoing. Indeed, several experiments have measured
these 
\[
R_{D^{(*)}}=\frac{\Gamma(B\to D^{(*)}\tau\nu_{\tau})}{\Gamma(B\to D^{(*)}\ell\nu_{\ell})},\;R_{K^{(*)}}=\frac{\Gamma(B\to K^{(*)}\tau^{+}\tau^{-})}{\Gamma(B\to K^{(*)}\ell^{+}\ell^{-})},\;\ell=e,\mu,
\]
and other semileptonic ratios and have seen deviations from the Standard
Model (SM), e.g. \cite{BaBar:2012obs,LHCb:2014vgu}. Nevertheless,
with time passing, new measurement arrived where these deviations
seem to progressively shrink \cite{LHCb:2023zxo,LHCb:2022vje,LHCb:2023uiv,Belle:2019rba,LHCb:2021trn}
(for \cite{LHCb:2021trn} see addendum). To shed an additional light
on this topic the leptonic decays of heavy quarkonia come handy. As
it was under quite general assumptions analyzed in \cite{Faroughy:2016osc},
a new physics (NP) contribution to the $b\to c\tau\nu$ transition
ought to influence also the $c\bar{c}\to\tau^{+}\tau^{-}$ and $b\bar{b}\to\tau^{+}\tau^{-}$
processes from where the motivation to study the latter.

The analysis of the leptonic decays of the $J/\psi$, $\Psi(2s)$
and $Y\left(\text{n}s\right)$ states has been addressed in numerous
works. Already decades ago, the quark model description of various
hadronic decays was given in \cite{VanRoyen:1967nq}, where the often
cited Van Royen-Weisskopf formula was presented: it predicts the electromagnetic
decay width $\Gamma_{\Upsilon\to\ell\ell}^{\text{EM}}$ be proportional
to $\left(1+2x_{\ell}^{2}\right)\left(1+4x_{\ell}^{2}\right)^{1/2}$,
$x_{\ell}=m_{\ell}/m_{\Upsilon}$. Significantly later, a highly cited
paper \cite{Eichten:1995ch} investigates quarkonium Schr\"{o}dinger
wave function at origin for several common potentials, which provides
essential inputs for evaluating various quarkonia observables. The
reference \cite{Sanchis-Lozano:2003xlk} already deals with the leptonic
universality breaking and proposes a light Higghs-like CP-odd particle
$A^{0}$ with a mass similar to $\Upsilon$ for the breaking to happen.
Coming back to potential models, one can mention \cite{Radford:2007vd},
which uses a semirelativistic Hamiltonian, linear confinement and
pQCD one-loop short distant potential to arrive to leptonic decay
widths which are, for the most part, in a good agreement with data.
Similarly, phenomenological potentials are studied in the non-relativistic
QCD approach to describe heavy quarkonia in \cite{Rai:2008sc}, and
again, leptonic decay widths are well described. Higher excited heavy
quarkonia states are, in an analogical way, analyzed in \cite{Shah:2012js}.
A nice review of charmonium states mentioning several theoretical
approaches is given in \cite{Voloshin:2007dx}. The author analyses
various properties of these states from the perspective of the underlying
theory of QCD, the first QCD correction to leptonic decay of the charmonium
in a non-relativistic picture is given in the Eq. (23) of the text.
An excellent investigation of possible NP contributions to $R_{D^{(*)}}$
using leptonic decays of quarkonia is, within the effective field
theory (EFT) approach, presented in \cite{Aloni:2017eny}. Here the
authors assume that the NP enters through gauge-invariant effective
operators of dimension six and that only the $\tau$ sector is affected.
They present a list of all relevant four-fermion operators, analyze
the decays (semileptonic and leptonic) in various NP scenarios and
show which of them are favored and which are not. A more recent publication
\cite{Wang:2019tqf} is based on the Bethe-Salpeter method. The authors
use a relativistic wave function combined with a simple kernel to
asses relativistic effects in the heavy quarkonia decays, which are,
as they claim, large. Many authors address short-distant corrections
to the leptonic decay widths of quarkonia which can be computed at
various precision order. This topic is covered in works \cite{Beneke:1997jm,Beneke:2014qea,Feng:2022vvk}
and references therein. An important branch of the theoretical investigation
is represented by the QCD on the lattice. The work \cite{Gray:2005ur}
uses non-relativistic QCD at NLO to describe $\Upsilon$ spectrum,
the electronic decay widths are presented too. A relativistic action
is used in \cite{Hatton:2020qhk}, where $J/\Psi$ is investigated
and its various parameters, including the width to two electrons,
are calculated, and also in \cite{Hatton:2021dvg}, where the authors
address the ground state $\Upsilon(1s)$ and compute also $\Gamma_{\Upsilon\to e^{+}e^{-}}$.

In this text we are interested in the leptonic decay widths of heavy
quarkonia in the framework of the covariant confined quark model (CCQM),
which is an EFT approach based on quark-hadron interaction Lagrangian.
The ground states $J/\Psi$ and $\Upsilon(1s)$ were already addressed
in \cite{Branz:2009cd}. Here, however, our research is extended in
two ways: we include into the consideration also higher excited states
and we apply a novel approach to describe these states. Also, the
here presented analysis is based on updated values of the model parameters
and thus numbers for ground states are modified too.

The text is organized as follows: we start by presenting very briefly
the CCQM, including references to our previous texts. Next we address
the transition amplitudes $V\to\gamma$, their expression within the
CCQM and their gauge invariance. By comparing the Lorentz structures
we identify and compute the appropriate form factors and predict decay
widths of ground states. Then we present the novel approach to describe
radial excitations: we implement quark mass running and introduce
the orthogonality requirement for the excited states. With this in
hand, we compute leptonic decay widths for higher vector states.

\section{CCQM and $V\to\gamma\to\ell\ell$ amplitude}

\subsection{Model}

The CCQM has been presented in great details in several works, see
e.g. \cite{Branz:2009cd,Ivanov:2006ni}. To keep the text short and
focus on the novel ideas, we give only a brief characteristic of the
model and provide the reader with references.
\begin{itemize}
\item The CCQM is based on an invariant quark-hadron non-local interaction
Lagrangian written in the case of mesons as
\begin{align*}
\mathcal{L}_{\text{int}} & =g_{M}M(x)J_{M}(x)+\text{H.c.},\\
J_{M}(x) & =\int dx_{1}\int dx_{2}\:F_{M}(x;x_{1},x_{2})\bar{q}_{2}(x_{2})\Gamma_{M}q_{1}(x_{1}),
\end{align*}
where $g_{M}$ is the coupling between the mesonic field $M$ and
the quark current $J_{M}$, $F_{M}$ is a vertex function (see hereunder)
and $\Gamma_{M}$ the appropriate string of Dirac matrices which corresponds
to the spin of the meson. The vertex function has two components:
a delta function which matches the barycenter of quarks to the meson
position and the remaining part, which is chosen to have an exponential
form in the momentum space mainly for computational reasons
\begin{align*}
F_{M}(x;x_{1},x_{2}) & =\delta(x-\omega_{1}x_{1}-\omega_{2}x_{2})\Phi_{M}\left[\left(x_{1}-x_{2}\right)^{2}\right],\\
\omega_{1,2} & =m_{1,2}/\left(m_{1}+m_{2}\right),\\
\Phi_{M}\left[\left(x_{1}-x_{2}\right)^{2}\right] & =\int\frac{d^{4}k}{(2\pi)^{4}}e^{-ik(x_{1}-x_{2})}\tilde{\Phi}(-k^{2}),\\
\tilde{\Phi}(-k^{2}) & =e^{k^{2}/\Lambda_{M}^{2}}.
\end{align*}
Here $\Lambda_{M}$ is a free parameter of the model, $m_{i}$ are
quark masses and the minus sign in the argument of $\tilde{\Phi}$
indicates that the exponent becomes negative in the Euclidean region
so that the expression has an appropriate fall-off behavior without
ultraviolet divergences. Besides hadron-related parameters $\Lambda_{M}$
the model comprises as parameters constituent quark masses. In situations
where the decaying particle is very heavy we implement an infrared
confinement (\cite{Branz:2009cd,Ganbold:2014pua}) to avoid decays
to quarks, it depends on an additional parameter $\lambda_{\text{cutoff}}$.
We do not use this mechanism in the present work.
\item The CCQM includes both, hadronic and mesonic fields as elementary,
one therefore needs to address the question of the double counting.
To avoid the latter we use the so-called compositeness condition,
which is expressed in terms of the derivative of the meson mass operator
$\Pi_{M}^{'}$
\begin{equation}
Z_{M}=1-g_{M}^{2}\Pi_{M}^{'}(m_{M}^{2})=0,\label{eq:compCond}
\end{equation}
where $Z_{M}$ is the renormalization constant. The meaning is that
the overlap between the physical state and the corresponding bare
state represented by $Z_{M}$ is set to zero, i. e. the physical state
does not contain the bare state and can be therefore interpreted as
bound, for details see \cite{Branz:2009cd,Dubnicka:2015iwg}. As a
result quarks exist only as virtual particles which are responsible
for the interaction, i.e. an initial state meson fluctuates into a
quark pair prior to interaction and is possibly re-created from quarks
in the final state. The equality in (\ref{eq:compCond}) is reached
by tuning the value of $g_{M}$, which is in this way determined.
The model does not contain gluons, their action is effectively taken
into the account by the parameter-dependent vertex function.
\item The inclusion of the electromagnetic (EM) interaction into a theory
with a non-local Lagrangian is not trivial. For the free parts of
the Lagrangian the usual minimal substitution is applied, it defines
$\mathcal{L}_{\text{int}}^{\text{EM}_{1}}$. For the interaction part
an approach inspired by \cite{Terning:1991yt} is used, i.e. the quark
fields are multiplied by a gauge field exponential
\begin{equation}
q_{i}(x)\to e^{-ie_{q_{i}}I(x_{i},x,P)}q_{i}(x),\quad I(x_{i},x,P)=\int_{x}^{x_{i}}dz_{\mu}A^{\mu}(z),\label{eq:gaugeFieldExp}
\end{equation}
where $P$ is a path between $x_{i}$ and $x$. Since only derivatives
of $I$ appear in the perturbative expansion, the latter is path-independent
\[
\frac{\partial}{\partial x^{\mu}}I(x,y,P)=A_{\mu}(x).
\]
The corresponding Lagrangian terms $\mathcal{L}_{\text{int}}^{\text{EM}_{2}}$
are then generated by the expansion of the gauge exponential (\ref{eq:gaugeFieldExp})
in orders of $A^{\mu}$, see \cite{Branz:2009cd,Branz:2010pq}. In
a diagrammatic representation the term $\mathcal{L}_{\text{int}}^{\text{EM}_{1}}$
generates photon line attached to the quark line, the $\mathcal{L}_{\text{int}}^{\text{EM}_{2}}$
part corresponds to a photon attached to the non-local vertex.
\end{itemize}
With this defined, we evaluate the appropriate Feynman diagrams and
get observable predictions. We use the Schwinger representation of
the quark propagators and perform computations analytically with the
exception of the last step, which is the integration over the space
of Schwinger parameters, which we do numerically.

\subsection{Amplitude}

\begin{figure}
\begin{centering}
\includegraphics[width=0.8\textwidth]{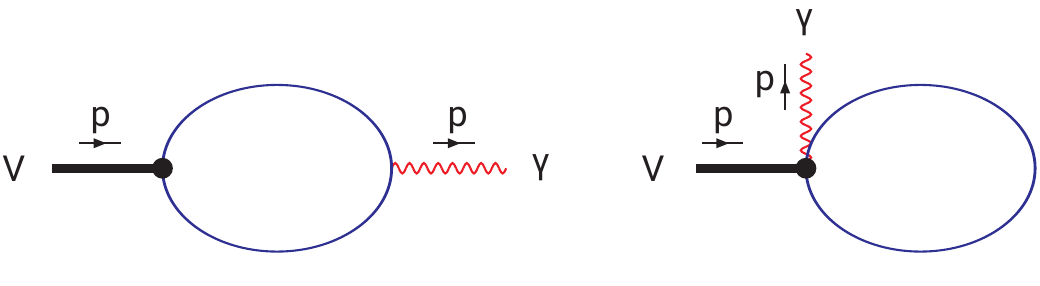}
\par\end{centering}
\caption{Diagrams describing the $V\to\gamma$ transition.}
\label{Fig:diagVG}

\end{figure}
The relevant quark model diagrams of the $V\to\gamma$ transition
are shown in Fig.~\ref{Fig:diagVG}. The corresponding Feynman one-loop
integrals read
\begin{align}
M^{\mu\nu}(p) & =N_{c}g_{V}\!\int\!\!\frac{d^{4}k}{(2\pi)^{4}i}\Big\{\Phi_{V}(-k^{2})\,\text{tr}\Big[\gamma^{\mu}S(k+\tfrac{1}{2}\,p)\gamma^{\nu}S(k-\tfrac{1}{2}\,p)\Big]\nonumber \\
 & \quad-\,\int\limits _{0}^{1}d\tau\Phi'_{V}(-z_{\tau})(2\,k+\tfrac{1}{2}\,p)^{\mu}\text{tr}\Big[\gamma^{\nu}S(k)\Big]\Big\},\label{eq:M_mu_nu}\\
z_{\tau} & =\tau(k+\tfrac{1}{2}\,p)^{2}+(1-\tau)k^{2}\,.\nonumber 
\end{align}
Here $N_{c}=3$ is the number of colors, $g_{V}$ is the coupling
of the vector meson with its constituent quarks, $S$ denotes the
quark propagator and $\Phi'_{V}$ is the derivative of the vertex
function with respect to its argument. In this paper we consider the
charmonium ($c\bar{c}$-state) and bottomonium ($b\bar{b}$-state).
The constituent quark propagator is written as
\[
S(k)=\frac{1}{m_{q}-\not\!k},
\]
where $m_{q}$ is the constituent quark mass. We use the Lorentz index
$\mu$ in relation to the photon field whereas $\nu$ corresponds
to the vector meson field. With the simple Gaussian form for the vertex
function one has
\begin{equation}
\Phi_{V}(-k^{2})=\exp(s\,k^{2}),\quad\Phi'_{V}(-k^{2})=-s\exp(s\,k^{2}),\quad\text{where}\quad s=1/\Lambda_{V}^{2}\,.\label{eq:vertexFunctionForm}
\end{equation}
Here the parameter $\Lambda_{V}$ characterizes the size of the vector
meson. One is allowed to use different choices for $\Phi_{V}$ as
long as it falls off sufficiently fast in the ultraviolet region of
Euclidean space to render the Feynman diagrams ultraviolet finite
\cite{Anikin:1995cf}.

\subsection{Gauge invariance}

As the first step in our study, let us prove that the amplitude $M^{\mu\nu}$
given by Eq.~(\ref{eq:M_mu_nu}) is gauge invariant, i.e. $M^{\mu\nu}p_{\mu}=0$.
The loop (or bubble) integral contracted with $p_{\mu}$ can be written
as
\begin{align}
 & \int\!\!\frac{d^{4}k}{(2\pi)^{4}i}\Phi_{V}\left(-k^{2}\right)\,\text{tr}\Big[\!\not\!p\,S(k+\tfrac{1}{2}\,p)\gamma^{\nu}S(k-\tfrac{1}{2}\,p)\Big]\nonumber \\
 & \quad=\int\!\!\frac{d^{4}k}{(2\pi)^{4}i}\Phi_{V}\left(-k^{2}\right)\,\Big\{\text{tr}\Big[S(k+\tfrac{1}{2}\,p)\gamma^{\nu}\Big]-\text{tr}\Big[S(k-\tfrac{1}{2}\,p)\gamma^{\nu}\Big]\Big\}\,.\label{eq:GI_1}
\end{align}
The integrand of the tadpole integral contracted with $p_{\mu}$ may
be transformed in the following way
\begin{align*}
 & -\int\limits _{0}^{1}d\tau\Phi'_{V}\Big(-\tau(k+\tfrac{1}{2}\,p)^{2}-(1-\tau)k^{2}\Big)\left(2\,kp+\tfrac{1}{2}\,p^{2}\right)\text{tr}\Big[\gamma^{\nu}S(k)\Big]\Big\}\\
 & \quad=-\frac{\Phi_{V}\Big(-(k+\tfrac{1}{2}\,p)^{2}\Big)-\Phi_{V}\Big(-k^{2}\Big)}{k^{2}-(k+\tfrac{1}{2}p)^{2}}\,2\,\left(kp+\tfrac{1}{4}\,p^{2}\right)\text{tr}\Big[\gamma^{\nu}S(k)\Big]\\
 & \quad=2\,\Big[\Phi_{V}\Big(-(k+\tfrac{1}{2}\,p)^{2}\Big)-\Phi_{V}\Big(-k^{2}\Big)\Big]\text{tr}\Big[\gamma^{\nu}S(k)\Big]\,.
\end{align*}
Finally, we use the identity
\[
\int d^{4}kf(k)\text{tr}\Big[\gamma^{\nu}S(k)\Big]=\frac{1}{2}\int d^{4}k\left(f(k)-f(-k)\right)\text{tr}\Big[\gamma^{\nu}S(k)\Big],
\]
which is valid for any well behaved function $f(k)$. We come to the
final expression for the tadpole integral contracted with the external
momentum $p$. One has
\begin{align}
 & -\int\!\!\frac{d^{4}k}{(2\pi)^{4}i}\int\limits _{0}^{1}d\tau\Phi'_{V}\Big(-\tau(k+\tfrac{1}{2}\,p)^{2}-(1-\tau)k^{2}\Big)\left(2\,kp+\tfrac{1}{2}\,p^{2}\right)\text{tr}\Big[\gamma^{\nu}S(k)\Big]\nonumber \\
 & \quad=\int\!\!\frac{d^{4}k}{(2\pi)^{4}i}\Big[\Phi_{V}\Big(-(k+\tfrac{1}{2}\,p)^{2}\Big)-\Phi_{V}\Big(-(k-\tfrac{1}{2}\,p)^{2}\Big)\Big]\text{tr}\Big[\gamma^{\nu}S(k)\Big]\nonumber \\
 & \quad=-\int\!\!\frac{d^{4}k}{(2\pi)^{4}i}\Phi_{V}\left(-k^{2}\right)\,\Big\{\text{tr}\Big[S(k+\tfrac{1}{2}\,p)\gamma^{\nu}\Big]-\text{tr}\Big[S(k-\tfrac{1}{2}\,p)\gamma^{\nu}\Big]\Big\}\,.\label{eq:GI_2}
\end{align}
It is readily seen that the expression given by Eq.~(\ref{eq:GI_1})
is exactly canceled out by the one given in Eq.~(\ref{eq:GI_2}).
Thus, we have $M^{\mu\nu}p_{\mu}=0$. It is interesting to see what
is going on in the case of a null momentum $p=0$. If $p=0$ then
the integral corresponding to the tadpole diagram may be transferred
to the one corresponding to the loop diagram by using the integration
by parts. One has
\begin{align}
 & \int\frac{d^{4}k}{(2\pi)^{4}i}\,\Big\{\Phi_{V}\left(-k^{2}\right)\text{tr}\Big[\gamma^{\mu}S(k)\gamma^{\nu}S(k)\Big]-2k^{\mu}\,\Phi'_{V}\left(-k^{2}\right)\text{tr}\Big[\gamma^{\nu}S(k)\Big]\Big\}\nonumber \\
 & \quad=\int\frac{d^{4}k}{(2\pi)^{4}i}\,\frac{\partial}{\partial k^{\mu}}\Big\{\Phi_{V}\left(-k^{2}\right)\text{tr}\Big[\gamma^{\nu}S(k)\Big]\Big\}=0.\label{eq:atPequalZero}
\end{align}
In the general case the invariant amplitude of the $V\to\gamma$ transition
is written as
\[
M^{\mu\nu}(p)=G_{V}(p^{2})\,g^{\mu\nu}+F_{V}(p^{2})\,p^{\mu}p^{\nu},
\]
where the form factor $G_{V}$ has a dimension of mass squared whereas
the form factor $F_{V}$ is dimensionless. The gauge invariance gives
the relation between the two form factors
\begin{equation}
G_{V}(p^{2})=-p^{2}F_{V}(p^{2})\label{eq:relation}
\end{equation}
valid for any $p^{2}$ value. From Eq.~(\ref{eq:atPequalZero}) follows
that both form factors do not have singularity at $p^{2}=0$. In the
case of decay $V\to\ell^{+}\ell^{-}$, only the first form factor
$G_{V}$ gives a contribution to the decay amplitude due to the conservation
of the vector lepton current.

\subsection{Results}

By using the standard technique of integration over the loop momentum
\emph{k} one can finalize the two-fold integrals in the expressions
for $V\to\gamma$ form factors. One can obtain the following expressions
for the integrals from the loop diagram
\begin{align*}
 & G_{V}^{{\rm \,loop}}(p^{2})=\\
 & \qquad\frac{3g_{V}}{(4\pi)^{2}}\!\int\limits _{0}^{\infty}\!\frac{dt\,t}{(s+t)^{2}}\int\limits _{0}^{1}\!d\alpha\,e^{z_{\alpha}}\Big\{4\,m_{q}^{2}+\frac{4}{s+t}+\Big[1-\frac{4\,t^{2}}{(s+t)^{2}}\left(\tfrac{1}{2}-\alpha\right)^{2}\Big]p^{2}\Big\},\\
 & F_{V}^{{\rm \,loop}}(p^{2})=\\
 & \qquad\frac{3g_{V}}{(4\pi)^{2}}\!\int\limits _{0}^{\infty}\!\frac{dt\,t}{(s+t)^{2}}\int\limits _{0}^{1}\!d\alpha\,e^{z_{\alpha}}\Big\{-2+\frac{8\,t^{2}}{(s+t)^{2}}\left(\tfrac{1}{2}-\alpha\right)^{2}\Big\},\\
 & z_{\alpha}=-t\left[m_{q}^{2}-\alpha(1-\alpha)\,p^{2}\right]+\frac{st}{s+t}\left(\tfrac{1}{2}-\alpha\right)^{2}p^{2}.
\end{align*}
In the case of the tadpole diagram, one has
\begin{align*}
G_{V}^{{\rm \,tad}}(p^{2}) & =\frac{3g_{V}}{(4\pi)^{2}}\!\int\limits _{0}^{\infty}\!\frac{dt\,s}{(s+t)^{2}}\int\limits _{0}^{1}\!d\tau\,e^{z_{\tau}}\Big[-\frac{4}{s+t}\Big],\\
F_{V}^{{\rm \,tad}}(p^{2}) & =\frac{3g_{V}}{(4\pi)^{2}}\!\int\limits _{0}^{\infty}\!\frac{dt\,s}{(s+t)^{2}}\int\limits _{0}^{1}\!d\tau\,e^{z_{\tau}}\Big[-\Big(1-\frac{2s\tau}{s+t}\Big)\frac{s\tau}{s+t}\Big],\\
z_{\tau} & =-tm_{q}^{2}+s\tau\left(1-\frac{s\tau}{s+t}\right)\frac{p^{2}}{4}.
\end{align*}
Then we define the full $V\to\gamma$ form factors as
\begin{align*}
G_{V}(p^{2}) & =G_{V}^{{\rm \,loop}}(p^{2})+G_{V}^{{\rm \,tad}}(p^{2})\,,\\
F_{V}(p^{2}) & =F_{V}^{{\rm \,loop}}(p^{2})+F_{V}^{{\rm \,tad}}(p^{2})\,.
\end{align*}
Their numerical values are calculated by using the FORTRAN codes which
require the NAG library.

The model parameters were determined by fitting calculated quantities
of basic processes to available experimental data or lattice simulations
(for details, see Ref.~\cite{Branz:2009cd}). Here, we take the values
of the constituent quark masses $m_{q}$ from an updated fit made
in \cite{Gutsche:2015mxa}. This fit improves the description of new
data on the exclusive $B$-meson and heavy baryon decays. Note that
in the fit, the infrared cutoff parameter $\lambda$ has been kept
fixed. The updated numerical values of the constituent quark masses
and the cutoff parameter $\lambda$ are given in Table~\ref{tab:fitmas}.
\begin{table}[H]
\begin{centering}
\begin{tabular}{cccccc}
\toprule 
$m_{u}$ & $m_{s}$ & $m_{c}$ & $m_{b}$ & $\lambda$ & \tabularnewline
$0.241$ & $0.428$ & $1.67$ & $5.04$ & $0.181$ & GeV\tabularnewline
\bottomrule
\end{tabular}
\par\end{centering}
\caption{Model parameters: quark masses and cutoff parameter $\lambda$ (all
in GeV).}
\label{tab:fitmas}

\end{table}
 Size parameters of charmonium and bottomonium are
\[
\Lambda_{J/\psi}=2.795(1)~{\rm GeV},\qquad\Lambda_{\Upsilon}=4.03(4)~{\rm GeV}.
\]
The masses of ground and radially excited states of charmonium and
bottomonium are given in Table~\ref{tab:mass}.
\begin{table}[H]
\begin{centering}
\begin{tabular}{cccccc}
\toprule 
$J/\Psi$ & $\Psi(2s)$ &  & $\Upsilon(1s)$ & $\Upsilon(2s)$ & $\Upsilon(3s)$\tabularnewline
$3096.900(6)$ & $3686.10(6)$ &  & $9460.30(26)$ & $10023.26(31)$ & $10355.2(5)$\tabularnewline
\bottomrule
\end{tabular}
\par\end{centering}
\caption{Masses of charmonia and bottomonia from PDG~\cite{ParticleDataGroup:2022pth}
(in MeV).}
\label{tab:mass}
\end{table}

First, we have checked numerically that the equation~(\ref{eq:relation})
is satisfied with high accuracy (6 digits calculated with relative
error $10^{-5}$). The term proportional to $p^{\mu}$ does not contribute
to the amplitude of the $V\to\gamma\to\ell^{-}\ell^{+}$ decay due
to the conservation of the lepton current. Therefore we will use the
form factor $G_{V}(p^{2})=-p^{2}F_{V}(p^{2})$ on the mass shell of
the vector meson $p^{2}=m_{V}^{2}$ for the calculation of the decay
width. One has
\[
\Gamma(V\to\ell^{+}\ell^{-})=\frac{4\pi\alpha^{2}}{3}\,m_{V}\Big(\frac{Q_{V}f_{V}}{m_{V}}\Big)^{2}\sqrt{1-4x_{\ell}^{2}}\,(1+2x_{\ell}^{2}),
\]
where $f_{V}\equiv G_{V}(m_{V}^{2})/m_{V}$, $x_{\ell}=m_{\ell}/m_{V}$
and the quark charge factor $Q_{V}=(e_{u}-e_{d})/\sqrt{2}=1/\sqrt{2}$
for the $\rho^{0}$ meson, $Q_{V}=e_{c}=2/3$ for $\psi$ mesons,
and $Q_{V}=e_{b}=-1/3$ for $\Upsilon$ mesons. Note that in our notation
the leptonic constant $f_{V}$ coincides with the weak leptonic constant
and has the dimension of mass. The results as predicted by the CCQM
are summarized in Table~\ref{Tab:results}. The theoretical errors
correspond to the propagated errors of hadronic parameters $\Lambda_{V}$,
which, on their turn, get the error from the fit of the CCQM to decay
width experimental data.
\begin{table}
\begin{centering}
\begin{tabular}{ccccc}
\toprule 
Quantity & \multicolumn{2}{c}{$J/\psi$} & \multicolumn{2}{c}{$\Upsilon(1s)$}\tabularnewline
$f_{V}$ (MeV) & \multicolumn{2}{c}{415.4} & \multicolumn{2}{c}{715}\tabularnewline
 & CCQM & Expt. & CCQM & Expt.\tabularnewline
$B(V\to\tau^{+}\tau^{-})$ (\%) &  &  & $2.46(7)$ & $2.60(10)$\tabularnewline
$B(V\to\mu^{+}\mu^{-})$ (\%) & $5.964(40)$ & $5.961(33)$ & $2.48(7)$ & $2.48(5)$\tabularnewline
$B(V\to e^{+}e^{-})$ (\%) & $5.964(40)$ & $5.971(32)$ & $2.48(7)$ & $2.38(11)$\tabularnewline
\bottomrule
\end{tabular}
\par\end{centering}
\caption{Leptonic branching fractions of $J/\psi$ and $\Upsilon(1s)$ compared
to experimental numbers.}
\label{Tab:results}

\end{table}

\section{New approach to radial excitations}

\subsection{First step: running quark mass in a loop}

We assume that the binding energy of the radial excitations of a quarkonium
$q\bar{q}$ is equal to the binding energy $E$ of the ground state
$V_{0}$: $E=m_{V_{0}}-2m_{q_{0}}$. This means that the value of
the constituent quark mass is for an excited state $V'$ equal to
\begin{equation}
m_{q'}=(m_{V'}-E)/2.\label{eq:quarkMasses}
\end{equation}
The ground state quark masses are taken from Table \ref{tab:fitmas}.
We find the values of the size parameters $\Lambda_{V}$ by fitting
the model to the experimental data on the leptonic decays $V\to\ell^{+}\ell^{-}$.
The size parameters and branching fractions predicted under this assumption
are shown in Tabs. \ref{Tab:runMassCharmonia} and \ref{Tab:runMassBottomonia}.
\begin{table}
\begin{centering}
\begin{tabular}{ccccc}
\toprule 
 & \multicolumn{2}{c}{$J/\psi$} & \multicolumn{2}{c}{$\Psi(2s)$}\tabularnewline
$\Lambda_{V}$(GeV) & \multicolumn{2}{c}{2.795} & \multicolumn{2}{c}{1.68}\tabularnewline
 & CCQM & Expt. & CCQM & Expt.\tabularnewline
$\mathcal{B}(V\to\tau^{+}\tau^{-})$(\%) & - &  & 0.31(1) & 0.31(4)\tabularnewline
$\mathcal{B}(V\to\mu^{+}\mu^{-})$(\%) & 5.964(40) & 5.961(33) & 0.80(2) & 0.80(6)\tabularnewline
$\mathcal{B}(V\to e^{+}e^{-})$(\%) & 5.964(40) & 5.971(32) & 0.80(2) & 0.79(2)\tabularnewline
\bottomrule
\end{tabular}
\par\end{centering}
\caption{Charmonia: size parameters and leptonic branching fractions with running
constituent quark masses.}
\label{Tab:runMassCharmonia}
\end{table}
\begin{table}
\begin{centering}
\begin{tabular}{ccccccc}
\toprule 
 & \multicolumn{2}{c}{$\Upsilon(1s)$} & \multicolumn{2}{c}{$\Upsilon(2s)$} & \multicolumn{2}{c}{$\Upsilon(3s)$}\tabularnewline
$\Lambda_{V}$(GeV) & \multicolumn{2}{c}{4.03} & \multicolumn{2}{c}{2.80} & \multicolumn{2}{c}{2.485}\tabularnewline
 & CCQM & Expt. & CCQM & Expt. & CCQM & Expt.\tabularnewline
$\mathcal{B}(V\to\tau^{+}\tau^{-})$(\%) & 2.46(7) & 2.60(10) & 1.90(0.5) & 2.00(21) & 2.17(3) & 2.29(30)\tabularnewline
$\mathcal{B}(V\to\mu^{+}\mu^{-})$(\%) & 2.48(7) & 2.48(5) & 1.92(0.5) & 1.93(17) & 2.18(3) & 2.18(21)\tabularnewline
$\mathcal{B}(V\to e^{+}e^{-})$(\%) & 2.48(7) & 2.38(11) & 1.92(0.5) & 1.91(16) & 2.18(3) & 2.18(20)\tabularnewline
\bottomrule
\end{tabular}
\par\end{centering}
\caption{Bottomonia: size parameters and leptonic branching fractions with
running constituent quark masses.}
\label{Tab:runMassBottomonia}
\end{table}

\subsection{Second step: orthogonality of the radial excitations}

\begin{figure}
\begin{centering}
\includegraphics[width=0.48\textwidth]{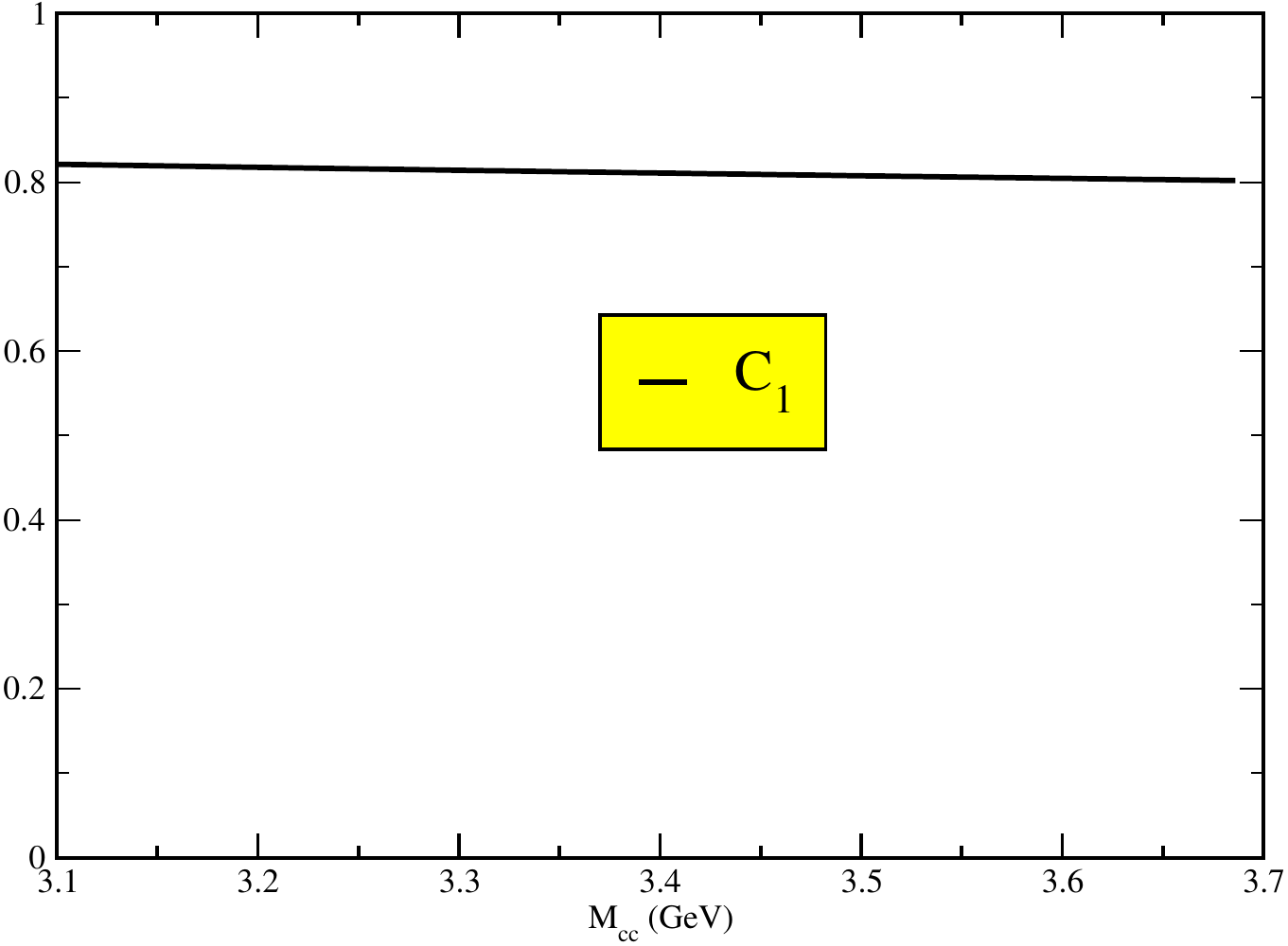} \includegraphics[width=0.48\textwidth]{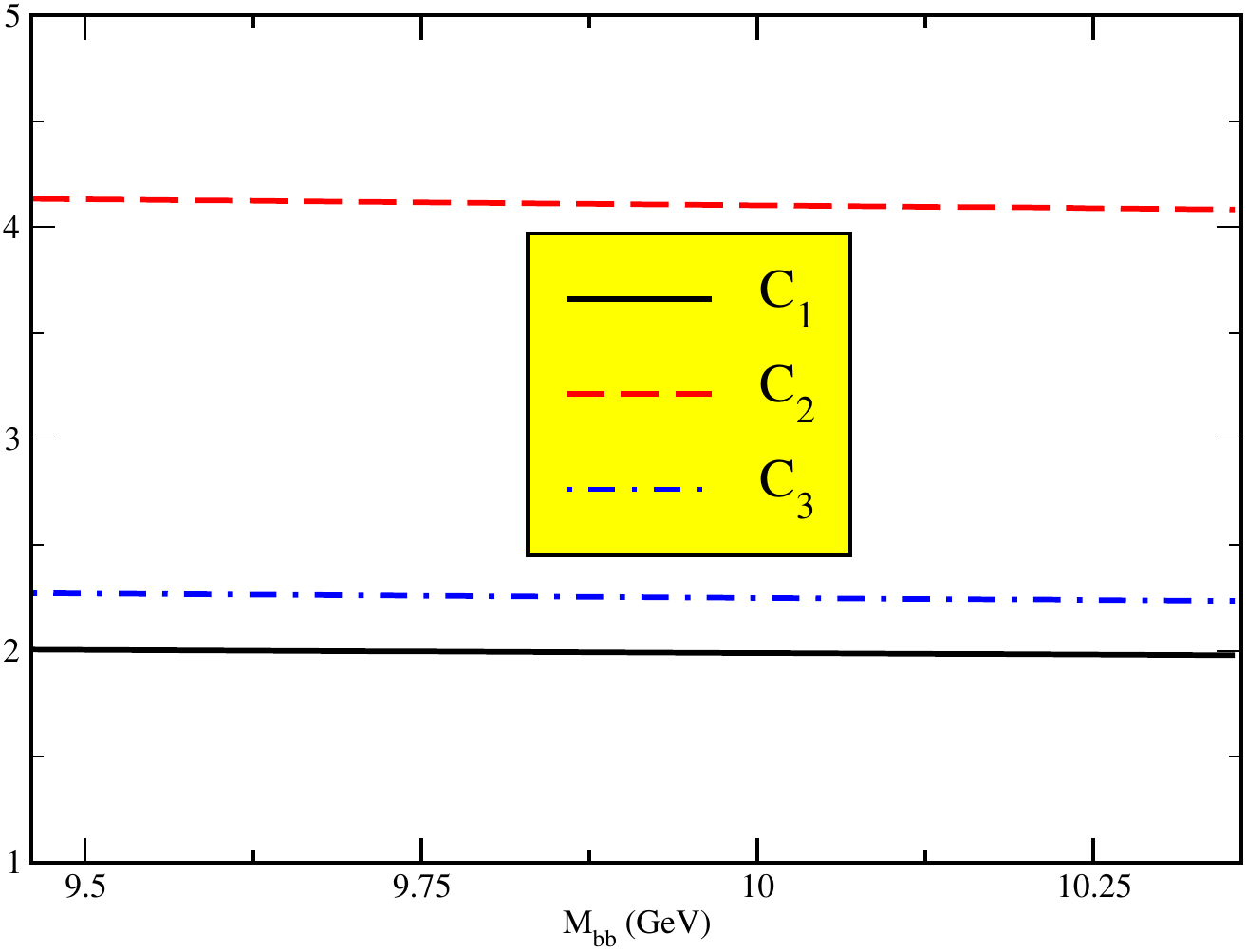}
\par\end{centering}
\caption{The dependence of coefficients of the orthogonal vertex functions
on the quarkonium mass for charmonia and bottomonia.}
\label{fig:c123_koefs}

\end{figure}

We assume that different radial excitations cannot pass into each
other through a quark loop. We call this assumption as the orthogonality
condition. To realize this condition we choose the vertex functions
in the following form
\begin{equation}
\Phi_{n}(-k^{2})=\Big(1+\sum\limits _{m=1}^{n}c_{m+n-1}s_{n}^{m}k^{2m}\Big)\Phi_{V}(-k^{2}),\label{eq:newVertex}
\end{equation}
where the function $\Phi_{V}(-k^{2})$ is given by Eq.~(\ref{eq:vertexFunctionForm}).
Then the orthogonality condition implies that
\begin{equation}
\left(g_{\mu\nu}-\frac{p_{\mu}p_{\nu}}{p^{2}}\right)\!\int\!\!\frac{d^{4}k}{(2\pi)^{4}i}\Phi_{n}(-k^{2})\,\Phi_{m}(-k^{2})\text{tr}\Big[\gamma^{\mu}S(k+\tfrac{1}{2}\,p)\gamma^{\nu}S(k-\tfrac{1}{2}\,p)\Big]=0.\label{eq:ortoghonality}
\end{equation}
This condition will allow us to determine the numerical values of
the coefficients $c_{m+n-1}$ in the allowed region of $p^{2}$ for
certain quarkonium spectrum. Here, we consider two radial excitations
of the charmonium, $J/\psi$ and $\Psi(2S)$, and three for the bottomonium:
$\Upsilon(1S)$, $\Upsilon(2S)$ and $\Upsilon(3S)$. We find the
values of coefficients $c_{i}$ which solve (\ref{eq:ortoghonality})
for charmonia and bottomonia separately, the solution depends on the
quarkonium mass. This dependence is shown in Fig.~\ref{fig:c123_koefs}.

\subsection{Results and discussion}

\begin{table}
\begin{centering}
\begin{tabular}{ccccc}
\toprule 
 & \multicolumn{2}{c}{$J/\psi$} & \multicolumn{2}{c}{$\Psi(2s)$}\tabularnewline
$\Lambda_{V}$(GeV) & \multicolumn{2}{c}{2.795} & \multicolumn{2}{c}{0.463}\tabularnewline
$c_{i}$ &  &  & \multicolumn{2}{c}{$c_{1}=0.802$}\tabularnewline
 & CCQM & Expt. & CCQM & Expt.\tabularnewline
$\mathcal{B}(V\to\tau^{+}\tau^{-})$(\%) & - &  & 0.31(1) & 0.31(4)\tabularnewline
$\mathcal{B}(V\to\mu^{+}\mu^{-})$(\%) & 5.964(40) & 5.961(33) & 0.81(2) & 0.80(6)\tabularnewline
$\mathcal{B}(V\to e^{+}e^{-})$(\%) & 5.964(40) & 5.971(32) & 0.81(2) & 0.79(2)\tabularnewline
\bottomrule
\end{tabular}
\par\end{centering}
\caption{Charmonia: size parameters and leptonic branching fractions in the
new approach with running constituent quark masses and orthogonal
vertex functions.}
\label{Tab:newApproachCharm}
\end{table}
\begin{table}
\begin{centering}
\begin{tabular}{ccccccc}
\toprule 
 & \multicolumn{2}{c}{$\Upsilon(1s)$} & \multicolumn{2}{c}{$\Upsilon(2s)$} & \multicolumn{2}{c}{$\Upsilon(3s)$}\tabularnewline
$\Lambda_{V}$(GeV) & \multicolumn{2}{c}{4.03} & \multicolumn{2}{c}{3.77} & \multicolumn{2}{c}{3.01}\tabularnewline
$c_{i}$ &  & \multicolumn{5}{c}{$c_{1}=1.979,\;c_{2}=4.083,\;c_{3}=2.235$}\tabularnewline
 & CCQM & Expt. & CCQM & Expt. & CCQM & Expt.\tabularnewline
$\mathcal{B}(V\to\tau^{+}\tau^{-})$(\%) & 2.46(7) & 2.60(10) & 1.92(0.5) & 2.00(21) & 2.17(3) & 2.29(30)\tabularnewline
$\mathcal{B}(V\to\mu^{+}\mu^{-})$(\%) & 2.48(7) & 2.48(5) & 1.93(0.5) & 1.93(17) & 2.18(3) & 2.18(21)\tabularnewline
$\mathcal{B}(V\to e^{+}e^{-})$(\%) & 2.48(7) & 2.38(11) & 1.93(0.5) & 1.91(16) & 2.18(3) & 2.18(20)\tabularnewline
\bottomrule
\end{tabular}
\par\end{centering}
\caption{Bottomonia: size parameters and leptonic branching fractions in the
new approach with running constituent quark masses and orthogonal
vertex functions.}
\label{Tab:newApproachBottom}
\end{table}
With quark masses defined by (\ref{eq:quarkMasses}) and modified
vertex functions (\ref{eq:newVertex}), where the coefficients are
determined by (\ref{eq:ortoghonality}), we settle, in an optimization
procedure, the value of size parameters $\Lambda_{V}$ so as to get
the best description of leptonic decay widths. We then get our results
summarized in Tabs. \ref{Tab:newApproachCharm} and \ref{Tab:newApproachBottom}.
The difference between the theory and the experimental values is,
in terms of standard deviations, shown in Table \ref{Tab:sigmas}.
\begin{table}
\begin{centering}
\begin{tabular}{ccccccc}
\toprule 
 & \multicolumn{6}{c}{$\left|\mathcal{B}_{\text{expt}}-\mathcal{B}_{\text{CCQM}}\right|/\sigma\left(\mathcal{B}_{\text{expt}}-\mathcal{B}_{\text{CCQM}}\right)$}\tabularnewline
 &  &  &  &  &  & \tabularnewline
 & $J/\psi$ & $\Psi(2s)$ &  & $\Upsilon(1s)$ & $\Upsilon(2s)$ & $\Upsilon(3s)$\tabularnewline
$\tau^{+}\tau^{-}$ &  & $\ll$1 &  & 1.15 & 0.38 & 0.40\tabularnewline
$\mu^{+}\mu^{-}$ & 0.06 & 0.16 &  & $\ll$1 & $\ll$1 & $\ll$1\tabularnewline
$e^{+}e^{-}$ & 0.12 & 0.71 &  & 0.77 & 0.12 & $\ll$1\tabularnewline
\bottomrule
\end{tabular}
\par\end{centering}
\caption{The difference between CCQM values and the experiment expressed in
terms of standard deviations.}
\label{Tab:sigmas}
\end{table}
 As readily seen, only one difference slightly exceeds $1\sigma$
deviation, others are within the $1\sigma$ interval. In this sense
we are able to describe the experimental measurements within our CCQM
SM approach. Taking into the account the relation with the semileptonic
decays mentioned in the Introduction, our results do not give support
for new physics phenomena being involved in the latter.

On the other hand, the results presented in Tabs. \ref{Tab:newApproachCharm}
and \ref{Tab:newApproachBottom} were reached as fits where parameters
were tuned and not as predictions. This is natural, since we are presenting
here a novel approach where the orthogonal vertex functions were presented
for the first time with their parameters not settled until now. It
is certainly desirable to analyze other processes in future, where
heavy quarkonia or possible other particles with their excited states
play a role. This will ensure that the number of parameters is significantly
smaller than the number of observables, i.e. the model will become
more constrained with more predictive power. Additionally, as part
of our outlook, the CCQM can be extended to include new physics EFT
operators, similarly to what has been presented in \cite{Aloni:2017eny}.
Then the corresponding Wilson coefficients can be analyzed for semileptonic
and leptonic decays and conclusions about their significance drawn
from the comparison to data.

\section*{Acknowledgment}

The research has been funded by the Science Committee of the Ministry
of Science and Higher Education of the Republic of Kazakhstan (Grant
No. AP19678771). Zh.T.'s work is supported by the JINR grant of young
scientists and specialists No. 24-301-06. The authors S.D., A.Z.D
and A.L. acknowledge the support of VEGA grant No. 2/0105/21.

\bibliographystyle{plunsrt}
\bibliography{quarkonia2leptons}

\begin{thebibliography}{10}

\bibitem{Kuhn:1998uy}
Johann~H. Kuhn, A.~A. Penin, A.~A. Pivovarov.
\newblock {Coulomb resummation for b anti-b system near threshold and precision
  determination of alpha(s) and m(b)}.
\newblock {\em Nucl. Phys. B}, 534:356--370, 1998.

\bibitem{Peset:2018ria}
Clara Peset, Antonio Pineda, Jorge Segovia.
\newblock {The charm/bottom quark mass from heavy quarkonium at N$^{3}$LO}.
\newblock {\em JHEP}, 09:167, 2018.

\bibitem{Resag:1994ki}
J.~Resag,  C.~R. Munz.
\newblock {Heavy quarkonia in the instantaneous Bethe-Salpeter model}.
\newblock {\em Nucl. Phys. A}, 590:735, 1995.

\bibitem{Narison:2018dcr}
Stephan Narison.
\newblock {QCD parameter correlations from heavy quarkonia}.
\newblock {\em Int. J. Mod. Phys. A}, 33(10):1850045, 2018.
\newblock [Addendum: Int.J.Mod.Phys.A 33, 1892004 (2018)].

\bibitem{QuarkoniumWorkingGroup:2004kpm}
N.~Brambilla,  i~in.
\newblock {Heavy quarkonium physics}.
\newblock 12 2004.

\bibitem{Brambilla:2010cs}
N.~Brambilla,  i~in.
\newblock {Heavy Quarkonium: Progress, Puzzles, and Opportunities}.
\newblock {\em Eur. Phys. J. C}, 71:1534, 2011.

\bibitem{Rothkopf:2019ipj}
Alexander Rothkopf.
\newblock {Heavy Quarkonium in Extreme Conditions}.
\newblock {\em Phys. Rept.}, 858:1--117, 2020.

\bibitem{BaBar:2012obs}
J.~P. Lees,  i~in.
\newblock {Evidence for an excess of $\bar{B} \to D^{(*)} \tau^-\bar{\nu}_\tau$
  decays}.
\newblock {\em Phys. Rev. Lett.}, 109:101802, 2012.

\bibitem{LHCb:2014vgu}
Roel Aaij,  i~in.
\newblock {Test of lepton universality using $B^{+}\rightarrow
  K^{+}\ell^{+}\ell^{-}$ decays}.
\newblock {\em Phys. Rev. Lett.}, 113:151601, 2014.

\bibitem{LHCb:2023zxo}
Roel Aaij,  i~in.
\newblock {Measurement of the ratios of branching fractions
  $\mathcal{R}(D^{*})$ and $\mathcal{R}(D^{0})$}.
\newblock {\em Phys. Rev. Lett.}, 131:111802, 2023.

\bibitem{LHCb:2022vje}
R.~Aaij,  i~in.
\newblock {Measurement of lepton universality parameters in $B^+\to
  K^+\ell^+\ell^-$ and $B^0\to K^{*0}\ell^+\ell^-$ decays}.
\newblock {\em Phys. Rev. D}, 108(3):032002, 2023.

\bibitem{LHCb:2023uiv}
Roel Aaij,  i~in.
\newblock {Test of lepton flavor universality using $B^0 \to D^{*-}
  \tau+\nu\tau$ decays with hadronic $\tau$ channels}.
\newblock {\em Phys. Rev. D}, 108(1):012018, 2023.

\bibitem{Belle:2019rba}
G.~Caria,  i~in.
\newblock {Measurement of $\mathcal{R}(D)$ and $\mathcal{R}(D^*)$ with a
  semileptonic tagging method}.
\newblock {\em Phys. Rev. Lett.}, 124(16):161803, 2020.

\bibitem{LHCb:2021trn}
Roel Aaij,  i~in.
\newblock {Test of lepton universality in beauty-quark decays}.
\newblock {\em Nature Phys.}, 18(3):277--282, 2022.
\newblock [Addendum: Nature Phys. 19, (2023)].

\bibitem{Faroughy:2016osc}
Darius~A. Faroughy, Admir Greljo, Jernej~F. Kamenik.
\newblock {Confronting lepton flavor universality violation in B decays with
  high-$p_T$ tau lepton searches at LHC}.
\newblock {\em Phys. Lett. B}, 764:126--134, 2017.

\bibitem{VanRoyen:1967nq}
R.~Van~Royen,  V.~F. Weisskopf.
\newblock {Hadron Decay Processes and the Quark Model}.
\newblock {\em Nuovo Cim. A}, 50:617--645, 1967.
\newblock [Erratum: Nuovo Cim.A 51, 583 (1967)].

\bibitem{Eichten:1995ch}
Estia~J. Eichten,  Chris Quigg.
\newblock {Quarkonium wave functions at the origin}.
\newblock {\em Phys. Rev. D}, 52:1726--1728, 1995.

\bibitem{Sanchis-Lozano:2003xlk}
Miguel~Angel Sanchis-Lozano.
\newblock {Leptonic universality breaking in upsilon decays as a probe of new
  physics}.
\newblock {\em Int. J. Mod. Phys. A}, 19:2183, 2004.

\bibitem{Radford:2007vd}
Stanley~F. Radford,  Wayne~W. Repko.
\newblock {Potential model calculations and predictions for heavy quarkonium}.
\newblock {\em Phys. Rev. D}, 75:074031, 2007.

\bibitem{Rai:2008sc}
Ajay~Kumar Rai, Bhavin Patel, P.~C. Vinodkumar.
\newblock {Properties of $Q \bar{Q}$ mesons in non-relativistic QCD formalism}.
\newblock {\em Phys. Rev. C}, 78:055202, 2008.

\bibitem{Shah:2012js}
Manan Shah, Arpit Parmar, P.~C. Vinodkumar.
\newblock {Leptonic and Digamma decay Properties of S-wave quarkonia states}.
\newblock {\em Phys. Rev. D}, 86:034015, 2012.

\bibitem{Voloshin:2007dx}
M.~B. Voloshin.
\newblock {Charmonium}.
\newblock {\em Prog. Part. Nucl. Phys.}, 61:455--511, 2008.

\bibitem{Aloni:2017eny}
Daniel Aloni, Aielet Efrati, Yuval Grossman, Yosef Nir.
\newblock {$\Upsilon$ and $\psi$ leptonic decays as probes of solutions to the
  $R_D^{(*)}$ puzzle}.
\newblock {\em JHEP}, 06:019, 2017.

\bibitem{Wang:2019tqf}
Guo-Li Wang,  Xing-Gang Wu.
\newblock {Revisiting the Heavy Vector Quarkonium Leptonic Widths}.
\newblock {\em Chin. Phys. C}, 44(6):063104, 2020.

\bibitem{Beneke:1997jm}
M.~Beneke, A.~Signer, Vladimir~A. Smirnov.
\newblock {Two loop correction to the leptonic decay of quarkonium}.
\newblock {\em Phys. Rev. Lett.}, 80:2535--2538, 1998.

\bibitem{Beneke:2014qea}
Martin Beneke, Yuichiro Kiyo, Peter Marquard, Alexander Penin, Jan Piclum, Dirk
  Seidel, Matthias Steinhauser.
\newblock {Leptonic decay of the $\Upsilon$(1$S$) meson at third order in QCD}.
\newblock {\em Phys. Rev. Lett.}, 112(15):151801, 2014.

\bibitem{Feng:2022vvk}
Feng Feng, Yu~Jia, Zhewen Mo, Jichen Pan, Wen-Long Sang, Jia-Yue Zhang.
\newblock {Complete three-loop QCD corrections to leptonic width of vector
  quarkonium}.
\newblock 7 2022.

\bibitem{Gray:2005ur}
A.~Gray, I.~Allison, C.~T.~H. Davies, Emel Dalgic, G.~P. Lepage, J.~Shigemitsu,
  M.~Wingate.
\newblock {The Upsilon spectrum and m(b) from full lattice QCD}.
\newblock {\em Phys. Rev. D}, 72:094507, 2005.

\bibitem{Hatton:2020qhk}
D.~Hatton, C.~T.~H. Davies, B.~Galloway, J.~Koponen, G.~P. Lepage, A.~T. Lytle.
\newblock {Charmonium properties from lattice $QCD$+QED : Hyperfine splitting,
  $J/\psi$ leptonic width, charm quark mass, and $a^c_\mu$}.
\newblock {\em Phys. Rev. D}, 102(5):054511, 2020.

\bibitem{Hatton:2021dvg}
D.~Hatton, C.~T.~H. Davies, J.~Koponen, G.~P. Lepage, A.~T. Lytle.
\newblock {Bottomonium precision tests from full lattice QCD: Hyperfine
  splitting, $\Upsilon$ leptonic width, and b quark contribution to $e^+e^-
  \rightarrow$ hadrons}.
\newblock {\em Phys. Rev. D}, 103(5):054512, 2021.

\bibitem{Branz:2009cd}
Tanja Branz, Amand Faessler, Thomas Gutsche, Mikhail~A. Ivanov, Jurgen~G.
  Korner, Valery~E. Lyubovitskij.
\newblock {Relativistic constituent quark model with infrared confinement}.
\newblock {\em Phys. Rev. D}, 81:034010, 2010.

\bibitem{Ivanov:2006ni}
Mikhail~A. Ivanov, Jurgen~G. Korner, Pietro Santorelli.
\newblock {Exclusive semileptonic and nonleptonic decays of the $B_c$ meson}.
\newblock {\em Phys. Rev. D}, 73:054024, 2006.

\bibitem{Ganbold:2014pua}
Gurjav Ganbold, Thomas Gutsche, Mikhail~A. Ivanov, Valery~E. Lyubovitskij.
\newblock {On the meson mass spectrum in the covariant confined quark model}.
\newblock {\em J. Phys. G}, 42(7):075002, 2015.

\bibitem{Dubnicka:2015iwg}
Stanislav Dubni\v{c}ka, Anna~Z. Dubni\v{c}kov\'a, Nurgul Habyl, Mikhail~A.
  Ivanov, Andrej Liptaj, Guliya~S. Nurbakova.
\newblock {Decay $B\to K^\ast(\to K\pi) \ell^+ \ell^-$ in covariant quark
  model}.
\newblock {\em Few Body Syst.}, 57(2):121--143, 2016.

\bibitem{Terning:1991yt}
John Terning.
\newblock {Gauging nonlocal Lagrangians}.
\newblock {\em Phys. Rev. D}, 44(3):887--897, 1991.

\bibitem{Branz:2010pq}
Tanja Branz, Amand Faessler, Thomas Gutsche, Mikhail~A. Ivanov, Jurgen~G.
  Korner, Valery~E. Lyubovitskij, Bettina Oexl.
\newblock {Radiative decays of double heavy baryons in a relativistic
  constituent three--quark model including hyperfine mixing}.
\newblock {\em Phys. Rev. D}, 81:114036, 2010.

\bibitem{Anikin:1995cf}
I.~V. Anikin, Mikhail~A. Ivanov, N.~B. Kulimanova, Valery~E. Lyubovitskij.
\newblock {The Extended Nambu-Jona-Lasinio model with separable interaction:
  Low-energy pion physics and pion nucleon form-factor}.
\newblock {\em Z. Phys. C}, 65:681--690, 1995.

\bibitem{Gutsche:2015mxa}
Thomas Gutsche, Mikhail~A. Ivanov, J\"urgen~G. K\"orner, Valery~E.
  Lyubovitskij, Pietro Santorelli, Nurgul Habyl.
\newblock {Semileptonic decay $\Lambda_b \to \Lambda_c + \tau^- +
  \bar{\nu_\tau}$ in the covariant confined quark model}.
\newblock {\em Phys. Rev. D}, 91(7):074001, 2015.
\newblock [Erratum: Phys.Rev.D 91, 119907 (2015)].

\bibitem{ParticleDataGroup:2022pth}
R.~L. Workman,  i~in.
\newblock {Review of Particle Physics}.
\newblock {\em PTEP}, 2022:083C01, 2022.

\end{thebibliography}

\end{document}